# A plan quality control method of treatment planning for Gamma Knife radiosurgery

Tonghe Wang, Matt D. Giles, Elizabeth Butker, Matthew C. Walb, Xiaofeng Yang, Tian Liu, Shannon Kahn, Zhen Tian

Department of Radiation Oncology, Emory University, Atlanta, GA30022

Email: zhen.tian@emory.edu

With many variables to adjust, conventional manual forward planning for Gamma Knife (GK) radiosurgery is very complicated and cumbersome. The resulting plan quality heavily depends on planners' skills, experiences and devoted efforts, and varies significantly among cases, planners, and institutions. Quality control for GK planning is desired to consistently provide high-quality plan to each patient. In this study, we proposed a quality control method for GK planning by building a database of high-quality GK plans. Patient anatomy was described by target volume, target shape complexity, and spatial relationship between target and nearby organs, which determine GK planning difficulty level. Plan quality was evaluated using target coverage, selectivity, intermediate dose spillage, maximum dose to 0.1 cc of brainstem, mean dose of ipsilateral cochlea, and beam-on time. When a new plan is created, a high-quality plan that has the most similar target volume size and shape complexity will be identified from the database. A model has also been built to predict the dose to brainstem and cochlea based on their overlap volume histograms. The identified reference plan and the predicted organ dose will help planners to make quality control decisions accordingly. To validate this method, we have built a database for vestibular schwannoma, which are considered to be challenging for GK planning due to the irregularly-shaped target and its proximity to brainstem and cochlea. Five cases were tested, among which one case was considered to be of high quality and four cases had a lower plan quality than prediction. These four cases were replanned and got substantially improved. Our results have demonstrated the efficacy of our proposed quality control method. This method may also be used as a plan quality prediction method to facilitate the development of automatic treatment planning for GK radiosurgery.

# 1 INTRODUCTION

Gamma Knife (GK) radiosurgery is an important and safe alternative to traditional neurosurgery for a variety of brain disorders, such as brain tumors [1,2], arteriovenous malformations [3,4], vestibular schwannomas [5-7], and meningiomas [8-10]. By focusing 192 narrow cobalt-60 beams on a single point, GK radiosurgery can eradicate the target with submillimeter accuracy while providing a rapid dose fall-off to spare the surrounding normal tissues [11].

An optimal treatment plan should be created for every individual patient to ensure the best treatment efficacy. Currently, the most commonly used treatment planning approach for GK radiosurgery is manual forward planning. Planners manually determine the number and locations of the radiation shots, beam-on time and collimator sizes of the eight physical sectors for each shot, and adjust these parameters via a trial-and-error schema to achieve a good treatment plan. This manual forward planning is very cumbersome and challenging, due to the large number of variables to adjust and the vast degrees of freedom involved. For instance, as the eight sectors can be independently driven to one of three different collimator sizes or be blocked entirely, there are $4^8$ (i.e., 65,536) available beam shapes to be selected for each shot. Adding to this complexity, any location within the target volume can be a potential shot location. A target of large volume size will substantially increase the degrees of freedom and the complexity of treatment planning. With such large degrees of freedom involved in treatment planning, it is impossible for planners to manually explore the entire solution space to search for the best plan for each specific patient [12-16]. Therefore, the resulting plan quality relies heavily on planners' skills, experiences, and the effort devoted to planning, and can vary significantly from case to case, planner to planner and institution to institution[17].

Unlike the medical linear accelerator (linac) based radiotherapy, the dosimetry of GK radiosurgery is significantly affected by the target size and target shape complexity due to the mechanical design of GK treatment unit [18]. The best achievable plan quality for each patient is essentially determined by his or her specific geometry. Therefore, although a few metrics, such as target coverage, selectivity, gradient index, dose delivered to organs at risk (OARs), and total beam-on time (BOT), are often used for GK plan evaluation, it is very difficult to use these metrics to directly determine whether a plan is of good plan quality for a specific patient anatomy. For instance, a target with an irregular shape is prone to result in less conformal dose distribution compared to a target with a spherical shape. Equivalent dose conformity achieved in a treatment plan doesn't necessarily translate to the same level of plan quality for different patient anatomy. Therefore, although provided with the values of these plan metrics, physicians still need to rely on their clinial judgement to decide whether to approve the current plan for treatment or try to further improve the plan quality. Plan quality control should be an essential component for GK radiosurgery to identify sub-optimal plans in order to ensure a consistently high plan quality for each individual patient. Yet, to our knowledge there is no plan quality control method available for GK radiosurgery, although significant effort has been devoted to developing quality control tools for linac-based radiotherapy [19-22].

In this study, we attempt to address this issue by building a database of high-quality GK treatment plans of the same disorder type and indexing these plans by target volume size, target shape complexity and the spatial relationship between the target and OARs. When a new treatment plan is created, a high-quality database plan that has the closest similarity to these anatomical parameters will be identified to serve as a reference for planners to estimate the expected plan quality and make plan quality control decision accordingly. In this study, we have built a database and tested its efficacy for vestibular schwannoma cases, which are usually considered to be challenging for GK planning due to the irregularly-shaped target and its proximity to brainstem and ipsilateral cochlea.

## 2 METHODS AND MATERIALS

*2.1 Patient database*

In this study, we have built a database of 22 previous patient cases, treated for vestibular schwannoma with GK radiosurgery with a Leksell GK ICON™ treatment unit (Elekta Instrument AB Stockholm, Stockholm, Sweden) at our institution from 2017-2019. All patients received a single fraction of 12.5 Gy, prescribed to 50% isodose level to allow higher dose within the tumor itself. For each patient, the original treatment plan was developed by our GK medical physicists using Leksell Gamma Plan (LGP) treatment planning system via manual forward planning, and reviewed and approved for treatment by our radiation oncologists and neurosurgeons. In order to build a database of high-quality GK plans for quality control purposes, after collecting and anonymizing these previous 22 clinical cases with IRB approval, two of our GK physicists replanned each case, trying to improve the original plan quality if possible. The same planning guidelines used in our institution to guide manual forward planning for GK radiosurgery of vestibular schwannoma cases were adopted to guide the re-planning as well: (1) 100% of prescription dose must be received by at least 99% of the target volume; (2) The maximum dose to 0.1 cc of brainstem must not exceed 12 Gy; (3) Try to keep the mean dose of the ipsilateral cochlea below 4 Gy; (4) Try to maximize selectivity and minimize gradient index to spare the nearby normal tissues as much as possible. For each case, the best plan among the two new plans and the original plan was selected to construct the database. The plan variables, 3D dose distribution, MR images, as well as the contours of target, brainstem, ipsilateral cochlea and skull, were included in our database. In addition, our database contains plan quality metrics and anatomic parameters for each case.

*2.1.1 Plan quality metrics*

In our database, we used six metrics for plan quality evaluation, i.e., coverage, selectivity, conformal index at 50% isodose line (denoted as CI50), maximum dose to 0.1 cc of brainstem (denoted as $D_{BS,0.1cc}$), mean dose of ipsilateral cochlea (denoted as $D_{CO,mean}$), and BOT normalized to a same dose rate. The coverage, selectivity and CI50 are defined as follows:

$$\text{coverage} = \frac{TV \cap PIV}{TV}. \tag{1}$$

$$\text{selectivity} = \frac{TV \cap PIV}{PIV}, \tag{2}$$

$$\text{CI50} = \frac{PIV_{R_x/2}}{TV}. \tag{3}$$

Here, TV and PIV represent the target volume and the planning isodose volume that receives at least the prescription dose, respectively. $PIV_{R_x/2}$ represents the planning isodose volume that receives at least half of the prescription dose. We would like to point out that we use CI50 instead of gradient index, which is defined as $PIV_{R_x/2}/PIV$ and used in LGP treatment planning system, because the gradient index was proposed to compare the dose gradient outside the target for plans of equal dose conformity and could not be used to directly compare plans of different conformity levels [23].

*2.1.2 Patient anatomical parameters*

As previously mentioned, the dosimetry of GK radiosurgery is affected by the target size and target shape complexity due to the mechanical design of GK treatment unit. The spatial target-OAR relationships (e.g., the proximity of target to brainstem and the ipsilateral cochlea for vestibular schwannoma cases) directly affect the amount of dose delivered to the OAR. Therefore, in our database we describe patient-specific anatomy in terms of target volume size, target shape complexity, and spatial target-OAR relationship to indicate the patient-specific difficulty level of GK planning.

Perimeter-area ratio (PARA) is the simplest shape index to measure the complexity of a 2D shape. It however varies with the shape size. For the same shape, an increase in size will cause a decrease in PARA. A new shape index, named SHAPE, was proposed by Patton to alleviate the size dependency issue of PARA by normalizing the PARA value to a standard shape of the same size [24]. Because of the circular shape of the collimators in GK treatment unit, we selected circle as the standard shape, and calculated the SHAPE value of the target contour on each image slice and used the mean value as our metric of target shape complexity. For demonstration purposes, Fig. 1 shows five 2D shapes and five target contours on the top and bottom rows, respectively, with their SHAPE values listed below.

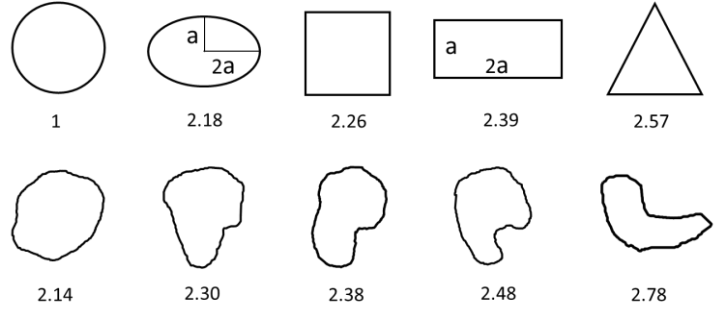

**Figure 1.** SHAPE values of five 2D shapes (top row) and five target contours (bottom row).

In our database, we employed the overlap volume histogram (OVH) to quantify the spatial relationship between target and brainstem (denoted as $OVH_{T\_BS}$) and that between target and ipsilateral cochlea (denoted as $OVH_{T\_CO}$) [19,25]. As illustrated in Fig. 2, OVH refers to the cumulative OAR volumes within a certain distance from the target surface, and an effective calculation way is to expand (or shrink) the target by the corresponding distance and calculate the OAR volume overlapped with the expanded (or shrinked) target. Please note that $OVH_{T\_BS}$ uses absolute volume corresponding to the dose limit of the maximum dose to 0.1cc brainstem, and $OVH_{T\_CO}$ uses percentage volume corresponding to the dose limit of the mean dose to cochlea. From these OVH curves, we have also obtained the target expansion that results in 0.1cc overlap volume for brainstem (denoted as $d_{BS}$, where $OVH_{T\_BS}(d_{BS}) = 0.1\ cc$) and the expansion that results in 50% overlap volume for cochlea ($d_{CO}$, where $OVH_{T\_CO}(d_{CO}) = 50\%$).

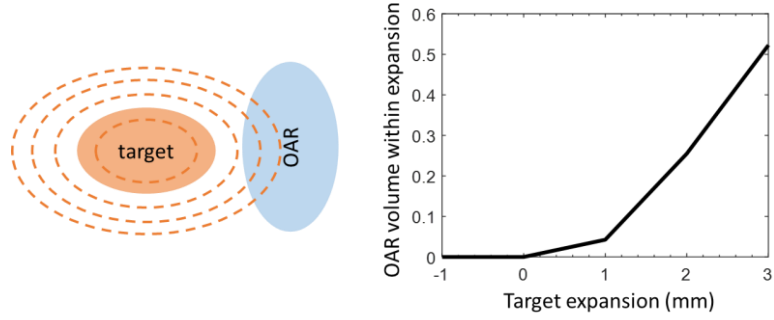

**Figure 2**. Illustration of overlap volume histogram.

*2.2 Quality control*

The details of our current database are listed in Table 1. When a treatment plan is created for a new patient case, ideally, the plan with the most similar anatomical parameters will be identified and retrieved from the database. This identified plan will serve as a high-quality reference plan for planners to estimate the expected plan quality and make plan quality control decision accordingly. However, due to the limited cases in our current database, for some test cases we cannot find a reference plan that has all the anatomical parameters very close to the test case. Therefore, based on our GK planning experiences, in this preliminary study we use target volume size and target SHAPE index as our criterion to select the reference case, and set up a threshold of maximum 10% target size difference. The scheme that we have designed to automatically select the reference plan is present in Table 2.

Table 1. Information of our initial patient database

| Disorder type | Vestibular schwannoma |
|---|---|
| Number of patient cases | 22 |
| Tumor (target) location | |
| • Number of cases on the left side | 11 |
| • Number of cases on the right side | 11 |
| Anatomic parameters | |
| • Target size | 0.093 cc – 2.779 cc |
| • SHAPE value (target shape complexity) | 2.14 – 2.69 |
| • Taget expansion to overlap with 0.1cc brainstem, that is, $d_{BS}$, where $OVH_{T\_BS}(d_{BS}) = 0.1 \, cc$ | 0.69 mm – 20.45 mm |
| • Target expansion to overlap with half volume of ispilateral cochlea, that is, $d_{CO}$, where $OVH_{T\_CO}(d_{CO}) = 50\%$ | 1.33 mm – 20.39 mm |
| Plan metrics | |
| • Coverage | 0.99 – 1.00 |
| • Selectivity | 0.55 – 0.87 |
| • CI50 | 3.11 – 6.04 |
| • $D_{BS,0.1cc}$ | 1.1 Gy – 12.0 Gy |
| • $D_{CO,mean}$ | 0.1 Gy – 9.9 Gy |
| • BOT normalized to a dose rate of 3.534 Gy/min* | 20.9 – 50.5 |

*3.534Gy/min is the initial dose rate when the radiation source was installed in our GK ICON$^{TM}$ unit and commissioned. BOT normalized to this initial dose rate is shorter than BOT of the actual treatment due to source decay.

Table 2. Our scheme to automatically identify the reference plan in the database

| Step 1. | For testing case, find in the database the cases that have target size within the 10% size difference threshold |
|---|---|
| | • If none within the threshold, no reference for the testing case |
| | • If only one case within the threshold, it will be selected as the reference |

Step 2. For those database cases that are within the size difference threshold, calculate the score as
$$s = \frac{|vol^t - vol^i|/vol^t}{1\%} + \frac{|SHAPE^t - SHAPE^i|}{0.01},$$
where $vol^t$ and $vol^i$ denote the volume sizes for the testing case and the ith candidate reference case, and $SHAPE^t$ and $SHAPE^i$ denote the SHAPE indices for the corresponding cases.

Step 3. Select the case with smallest score as the reference case.

- If more than one case, go to step 2.

With this scheme, the identified reference case may have a very different spatial target-OAR relationship, compared with the testing case. Hence, to roughly estimate the reasonable $D_{BS,0.1cc}$ and $D_{CO,mean}$ for the testing case, we have built a model between $d_{BS}$ and $D_{BS,0.1cc}$ and a model between $d_{CO}$ and $D_{CO,mean}$, which are formulated as

$$D_{BS,0.1cc}(d_{BS}) = a_1(d_{BS})^{b_1} + c_1, \qquad (3)$$

$$D_{CO,mean}(d_{CO}) = a_2(d_{CO})^{b_2} + c_2. \qquad (4)$$

The curve fitting toolbox ctfool in Matlab (The Mathworks, Inc.) was employed to fit the parameters in these two models, using the 22 cases in our current database. The models are shown in Fig. 3. We would like to mention that our fitted cochlea dose model may not be very reliable when $d_{CO}$ is bigger than 3.3 mm, due to the very few cases in our current database that have large $d_{CO}$. The OVH curves of the testing case and the identified reference case will also be provided to planners to help them make quality control decisions.

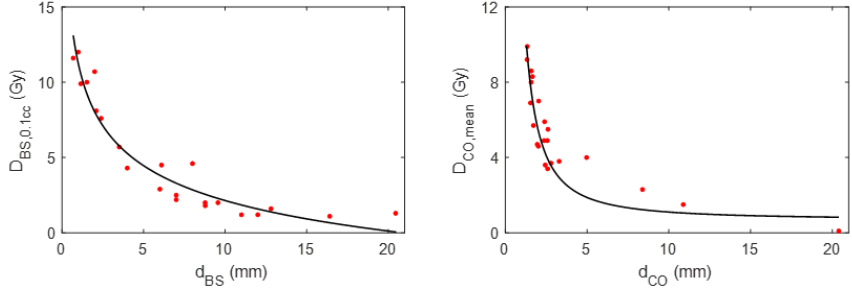

**Figure 3**. The models fitted using the database to estimate the reasonable branstem 0.1cc dose and cochlea mean dose, based on the target expansion distance $d_{BS}$ and $d_{CO}$ that results in 0.1cc brainstem overlap volume and 50% cochlea overlap volume, respectively.

## 3 RESULTS

In this preliminary study, we tested this plan quality control method on five cases using the built patient database. The information of the testing plans and the corresponding identified reference plans are listed in Table 3. The $OVH_{T\_BS}$ and $OVH_{T\_CO}$ that describe the spatial relationship between target and brainstem and between target and ipsilateral cochlea for each case are shown in the first two columns in Fig 4. The brainstem 0.1cc dose and cochlea mean dose obtained in each plan, as well as our fitted dose models, are shown in the last two columns in Fig 4.

**Table 3**. Information of the testing cases (Ti, i=1,2,…,5) and their corresponding reference cases (Ri, i=1,2,…,5) identified from the patient database.

|    | target size (cc) | SHAPE | $d_{BS}$ (mm) | $d_{CO}$ (mm) | coverage | selectivity | CI50 | BOT$_n$ (min) | $D_{BS,0.1cc}$ (Gy) | $D_{CO,mean}$ (Gy) |
|----|------|------|------|------|------|------|------|------|------|------|
| T1 | 0.777 | 2.34 | 9.10 | 2.28 | 1.00 | 0.69 | 4.01 | 61.0 | 2.8 | 4.2 |
| R1 | 0.790 | 2.64 | 2.40 | 1.58 | 1.00 | 0.71 | 3.94 | 43.0 | 7.6 | 8.0 |
| T2 | 0.289 | 2.46 | 6.00 | 2.49 | 1.00 | 0.56 | 5.29 | 41.2 | 2.5 | 3.7 |
| R2 | 0.303 | 2.45 | 9.56 | 1.54 | 1.00 | 0.71 | 4.18 | 31.9 | 2.0 | 6.9 |
| T3 | 0.283 | 2.59 | 10.64 | 2.24 | 0.99 | 0.43 | 7.05 | 42.4 | 1.8 | 4.0 |
| R3 | 0.284 | 2.56 | 8.78 | 2.56 | 0.99 | 0.70 | 5.30 | 23.7 | 1.8 | 3.4 |
| T4 | 2.784 | 2.20 | 1.20 | 3.60 | 1.00 | 0.79 | 3.20 | 52.4 | 10.4 | 5.1 |

| | | | | | | | | | |
|---|---|---|---|---|---|---|---|---|---|
| R4 | 2.779 | 2.14 | 1.52 | 4.97 | 1.00 | 0.81 | 3.30 | 37.1 | 10.0 | 4.0 |
| T5 | 0.415 | 2.62 | 3.25 | 2.99 | 1.00 | 0.67 | 4.73 | 39.7 | 5.5 | 4.0 |
| R5 | 0.418 | 2.54 | 7.00 | 2.37 | 1.00 | 0.65 | 4.66 | 36.5 | 2.5 | 4.9 |

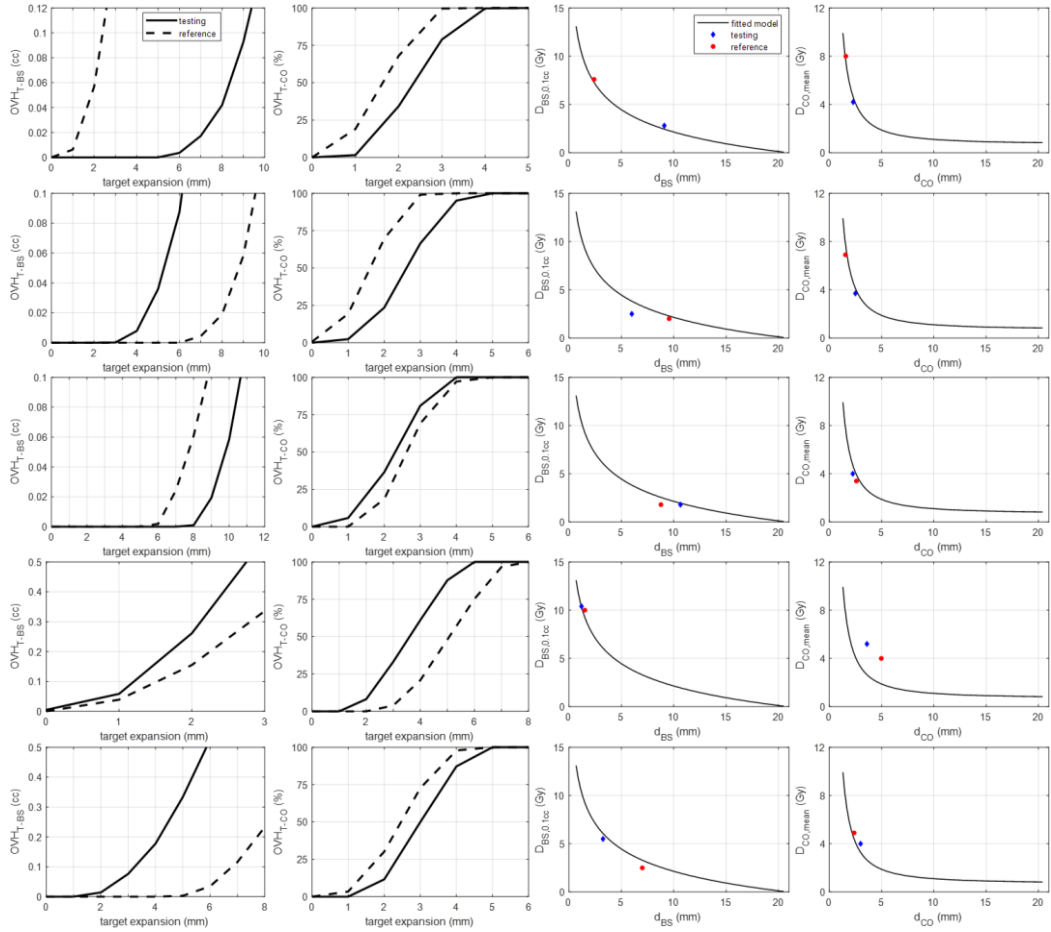

**Figure 4.** The five rows are corresponding to the five testing cases T1-T5, respectively. $OVH_{T\_BS}$ and $OVH_{T\_CO}$ are shown in the first two columns, with the OVH curves for the testing case in solid line and the curves for the reference case in dashed line. Our fitted dose model for $D_{BS,0.1cc}$ and $D_{CO,mean}$ are shown in the last two columns, and the doses obtained in the testing case and in the reference case are denoted by a diamond shape and a solid circle, respectively.

With the anatomical parameters and the plan metrics of the identified reference plans, planners can estimate the expected plan quality for the testing plans and make plan quality control decisions accordingly. For instance, the target in testing case T1 has similar volume size but much smaller SHAPE index, and is relatively further away from both brainstem and cochlea, compared to the target in the corresponding high-quality reference plan R1. The geometry in T1 is relatively easier for GK planning than the geometry in R1. Compared with the much higher selectivity (i.e., 0.71) and shorter BOT (i.e., 43.0 min) achieved in R1, replanning is suggested for the testing T1 case that has a selectivity of 0.69 and a BOT of 61.0 min in the current plan. As shown in Fig. 4, the OAR dose obtained in T1's current plan agree with our fitted model, and hence expecting no big improvement on them in replanning. With similar target sizes and SHAPE values in T2 and R2 cases and reasonable OAR doses obtained in both cases, significant quality improvement is expected for T2 to improve the selectivity, CI50, and BOT to be comparable to those in R2. Similarly, large improvement on selectivity, CI50, and BOT is also expected for T3. It is observed that the cochlea mean dose obtained in both T4 and R4 cases have a big deviation from our fitted dose model. This may be ascribed to the less reliability of our fitted model at large $d_{CO}$ values. Nevertheless, the relatively higher cochlea dose obtained in T4 testing plan is consistent with its smaller distance between target and cochlea. With the similar target size and SHAPE values in T4 and R4 cases and similar dose conformity obtained, large improvement on BOT is expected for T4. The testing T5 plan obtains similar values of coverage, selectivity, CI50 and BOT compared to the R5 plan, which is consistent with their similar target sizes and SHAPE values. As shown in Fig.4, OAR doses obtained in both cases are close to our built dose models. Hence, the T5 plan is considered to be at a high quality level, comparable to the quality of R5, and therefore doesn't necessitate replanning.

Our physicist replanned T1-4 cases according to the quality control results. The plan metrics of the obtained new plans are shown in Table 4. As expected by our quality control method, all the five replanned cases gained significant quality improvement. For all the cases except T2, the plan quality was improved at no cost of any of the plan metrics. For instance, better selectivity (0.74 vs. 0.69) and CI50 (3.82 vs. 4.01), shorter BOT (43.0 min vs. 61.0 min) and lower brainstem dose (2.6 Gy vs. 2.8 Gy) were obtained in replanning for T1 case while achieving the same coverage and cochlea mean dose. For T3 case, the new plan significantly improved selectivity (0.68 vs. 0.43), CI50 (4.82 vs. 7.05) and BOT (30.2 min vs. 42.4 min), while achieving slightly lower dose for brainstem and cochlea and same coverage. As shown in Fig 5, the dose conformity was improved in the new plan T3-re to a level comparable to that of reference plan R4. For case 4, BOT was substantially reduced in the new plan (from 52.4 min to 37.8 min), while achieving similar or slightly better values for other plan metrics. For T2 case, although the brainstem dose $D_{BS,0.1cc}$ was increased from 2.5 Gy to 3.4 Gy, it is still well below our 12 Gy dose limit. Meanwhile, both selectivity and BOT were improved substantially (i.e., from 0.56 and 41.2 min to 0.69 and 29.8 min).

**Table 4**. Plan metrics of the new plans obtained by replanning for T1-4 cases (denoted as Ti-re, i=1,2,…,4). The plan metrics of the original plans listed in Table 3 are also listed here for comparison.

|  | coverage | selectivity | CI50 | $BOT_n$ (min) | $D_{BS,0.1cc}$ (Gy) | $D_{CO,mean}$ (Gy) |
|---|---|---|---|---|---|---|
| T1 | 1.00 | 0.69 | 4.01 | 61.0 | 2.8 | 4.2 |
| T1-re | 1.00 | 0.74 | 3.82 | 43.0 | 2.6 | 4.2 |
| T2 | 1.00 | 0.56 | 5.29 | 41.2 | 2.5 | 3.7 |
| T2-re | 1.00 | 0.69 | 5.04 | 29.8 | 3.4 | 3.6 |
| T3 | 0.99 | 0.43 | 7.05 | 42.4 | 1.8 | 4.0 |
| T3-re | 0.99 | 0.68 | 4.82 | 30.2 | 1.3 | 3.6 |
| T4 | 1.00 | 0.79 | 3.20 | 52.4 | 10.4 | 5.2 |
| T4-re | 1.00 | 0.81 | 3.20 | 37.8 | 10.4 | 4.1 |

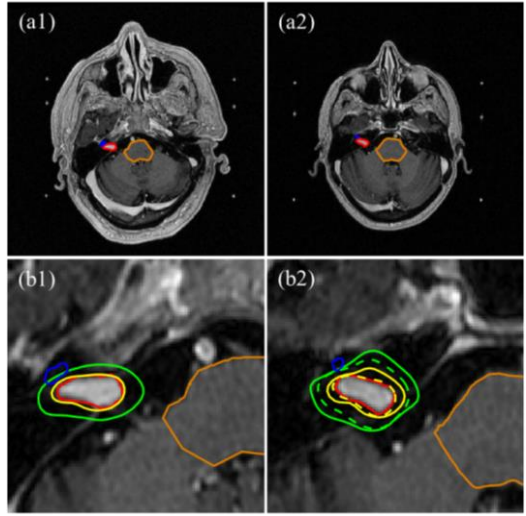

**Figure 5.** Replanning results of T3 case. (a1-2) shows a transverse slice of MR image for R3 and T3 cases, respectively. Contours of target, brainstem and ipsilateral cochlea are shown in red, orange and blue, respectively. Target regions of (a1-2) are zoomed in and shown in (b1-2), with the isodose lines of the prescription dose (in yellow) and half of the prescription dose (in green). In (b2), the isodose lines of the original T3 plan are shown in solid lines and those of the new plan T3-re are shown in dashed lines.

## 4 CONCLUSION AND DISCUSSION

Plan quality of GK treatment plans generated via manual forward planning can vary significantly from case to case, planner to planner and institution to institution. In order to consistently achieve high quality plans for each individual patient, we developed a novel quality control method for GK planning, and tested it for vestibular schwannoma cases. In this method, we have built a patient database of high-quality treatment plans, and used anatomical parameters, i.e., target volume size, SHAPE, $OVH_{T\_BS}$, $OVH_{T\_CO}$, as well as $d_{BS}$ and $d_{CO}$ obtained from OVH curves, to assess the planning difficulty level. When a treatment plan is created for a new patient case, a high-quality plan of similar planning difficulty level in the database is identified based on the values of the anatomy descriptors, and serves as a reference for planners to make quality control decisions accordingly. In this study, we have tested our method on five cases, among which four cases were found to have much worse plan quality than the corresponding reference plans and thus justified replanning. Distinct quality improvement has been achieved in these cases after replanning, demonstrating the efficacy of our quality control method.

 We would like to mention that we have attempted to extend the shape index from 2D to 3D in our experiments, that is, calculate the normalized surface area to volume ratio for target volume in relative to a sphere of the same volume size. However, it was found that this 3D shape index couldn't well represent the planning difficulty level compared to the 2D shape index. This might be explained by two reasons. First, we have found that surface area calculation was heavily dependent on the various smoothness level of the contours along superior-inferior direction, caused by the variation in slice thickness among different cases. Surface area calculation is also sensitive to the inter-slice contouring variability. Hence, it is difficult to fairly compare the obtained values of the normalized surface area to volume ratio among cases. In contrast, the SHAPE index value averaged over slices make itself less sensitive to these issues. Second, in manual forward planning, planners usually place shots every few slices in the transverse view, which might make 2D shape complexity more correlated to the resulting plan quality.

One limitation of this study is the small amount of the high-quality GK plans in the current database that we have built for vestibular schwannoma cases. This is because of the low incidence of vestibular schwannoma compared to other brain tumors as well as the patient population in our institution. Another reason is the considerable amount of human effort involved to build this database, as every reference case needs to be replanned to ensure the high plan quality. For this preliminary study, we started with this database of limited cases to test the efficacy and feasibility of our quality control method, and it will continue to be updated with new cases. Due to the limited cases in the current database, for some testing cases it is possible that we cannot find a reference plan with all the anatomy descriptors very close to the testing case. In this study, we used target volume size and SHAPE index as our primary criteria to select the reference case. In addition, we used $\boldsymbol{d_{BS}}$ and $\boldsymbol{d_{CO}}$ obtained from OVH curves to roughly estimate the OAR doses that can be obtained when achieving good dose conformity level and BOT, based on our OAR dose models. Currently, our fitted cochlea dose model may not be very reliable at $\boldsymbol{d_{CO}}$ larger than 3.3 mm, due to the very few cases with large $\boldsymbol{d_{CO}}$ in our current database. This issue is expected to be gradually relieved as new cases are added to the database to increase its diversity.

Another limitation of this study is that we can only estimate the achievable plan quality of a new case with a similar trade-off adopted in its corresponding reference case. We cannot predict its plan quality if a different trade-off is preferred for the new case. For instance, if the new patient has received an irradiation prior to this radiosurgery procedure and hence needs a much higher priority on reducing cochlea dose than improving target coverage and dose conformity, we cannot predict the plan quality at this situation. We will address this issue in our future work. One solution is to create a series of high-quality reference plans with different trade-offs. Another potential solution is to build a model among plan metrics for different anatomies to predict how much other plan metrics will be affected by improving one plan metric.

Currently in linac-based radiotherapy, a great deal of effort has been devoted towards automatic treatment planning, such as knowledge-based auto-planning [26-28] and deep-learning auto-planning [29,30]. An important step in auto-planning is plan quality prediction to predict achievable plan quality to guide the auto-planning [31-33]. Compared to linac-based radiotherapy, automatic treatment planning is even more desirable for GK radiosurgery to improve its planning efficiency, as GK radiosurgery is typically a one-day procedure and patients wait for the planning to be done while wearing a head frame. Our plan quality control method proposed in this study may also be used a plan quality prediction method for GK planning, facilitating the development of automatic treatment planning for GK radiosurgery.

**ACKNOWLEDGEMENTS**

This research is supported by Winship Cancer Institute #IRG-17-181-06 from the American Cancer Society.

**REFERENCES**

1. Muacevic A, Kreth FW, Horstmann GA, et al. Surgery and radiotherapy compared with gamma knife radiosurgery in the treatment of solitary single cerebral metastases of small diameter. *Journal of neurosurgery.* 1999;91(1):35-43.
2. Petrovich Z, Yu C, Giannotta SL, O'day S, Apuzzo ML. Survival and pattern of failure in brain metastasis treated with stereotactic gamma knife radiosurgery. *Journal of neurosurgery.* 2002;97(Supplement 5):499-506.
3. Karlsson B, Lindquist C, Steiner L. Prediction of obliteration after gamma knife surgery for cerebral arteriovenous malformations. *Neurosurgery.* 1997;40(3):425-431.
4. Schneider BF, Eberhard DA, Steiner LE. Histopathology of arteriovenous malformations after gamma knife radiosurgery. *Journal of neurosurgery.* 1997;87(3):352-357.
5. Lunsford LD, Niranjan A, Flickinger JC, Maitz A, Kondziolka D. Radiosurgery of vestibular schwannomas: summary of experience in 829 cases. *Journal of neurosurgery.* 2005;102(Special_Supplement):195-199.
6. Myrseth E, Møller P, Pedersen P-H, Vassbotn FS, Wentzel-Larsen T, Lund-Johansen M. Vestibular schwannomas: clinical results and quality of life after microsurgery or gamma knife radiosurgery. *Neurosurgery.* 2005;56(5):927-935.
7. Myrseth E, Møller P, Pedersen P-H, Lund-Johansen M. Vestibular schwannoma: surgery or gamma knife radiosurgery? A prospective, nonrandomized study. *Neurosurgery.* 2009;64(4):654-663.
8. Kondziolka D, Lunsford LD, Coffey RJ, Flickinger JC. Stereotactic radiosurgery of meningiomas. *Journal of neurosurgery.* 1991;74(4):552-559.
9. Iwai Y, Yamanaka K, Ishiguro T. Gamma knife radiosurgery for the treatment of cavernous sinus meningiomas. *Neurosurgery.* 2003;52(3):517-524.
10. Kreil W, Luggin J, Fuchs I, Weigl V, Eustacchio S, Papaefthymiou G. Long term experience of gamma knife radiosurgery for benign skull base meningiomas. *Journal of Neurology, Neurosurgery & Psychiatry.* 2005;76(10):1425-1430.
11. Wu A, Lindner G, Maitz A, et al. Physics of gamma knife approach on convergent beams in stereotactic radiosurgery. *International Journal of Radiation Oncology* Biology* Physics.* 1990;18(4):941-949.
12. Wu QJ, Chankong V, Jitprapaikulsarn S, et al. Real‐time inverse planning for Gamma Knife radiosurgery. *Medical physics.* 2003;30(11):2988-2995.
13. Levivier M, Carrillo RE, Charrier R, Martin A, Thiran J-P. A real-time optimal inverse planning for Gamma Knife radiosurgery by convex optimization: description of the system and first dosimetry data. *Journal of neurosurgery.* 2018;129(Suppl1):111-117.
14. Ghobadi K, Ghaffari HR, Aleman DM, Jaffray DA, Ruschin M. Automated treatment planning for a dedicated multi‐source intracranial radiosurgery treatment unit using projected gradient and grassfire algorithms. *Medical physics.* 2012;39(6Part1):3134-3141.
15. Sjölund J, Riad S, Hennix M, Nordström H. A linear programming approach to inverse planning in Gamma Knife radiosurgery. *Medical physics.* 2019;46(4):1533-1544.
16. Tian Z, Yang X, Giles M, et al. A preliminary study on a multiresolution‐level inverse planning approach for Gamma Knife radiosurgery. *Medical Physics.* 2020;47(4):1523-

1532.

17. Massager N, Lonneville S, Delbrouck C, Benmebarek N, Desmedt F, Devriendt D. Dosimetric and clinical analysis of spatial distribution of the radiation dose in gamma knife radiosurgery for vestibular schwannoma. *International Journal of Radiation Oncology* Biology* Physics.* 2011;81(4):e511-e518.
18. Borden JA, Mahajan A, Tsai J-S. A quality factor to compare the dosimetry of gamma knife radiosurgery and intensity-modulated radiation therapy quantitatively as a function of target volume and shape. *Journal of neurosurgery.* 2000;93(supplement_3):228-232.
19. Wu B, Ricchetti F, Sanguineti G, et al. Patient geometry‐driven information retrieval for IMRT treatment plan quality control. *Medical physics.* 2009;36(12):5497-5505.
20. Moore KL, Brame RS, Low DA, Mutic S. Experience-based quality control of clinical intensity-modulated radiotherapy planning. *International Journal of Radiation Oncology* Biology* Physics.* 2011;81(2):545-551.
21. Song T, Staub D, Chen M, et al. Patient-specific dosimetric endpoints based treatment plan quality control in radiotherapy. *Physics in Medicine & Biology.* 2015;60(21):8213.
22. Li N, Carmona R, Sirak I, et al. Highly efficient training, refinement, and validation of a knowledge-based planning quality-control system for radiation therapy clinical trials. *International Journal of Radiation Oncology* Biology* Physics.* 2017;97(1):164-172.
23. Paddick I, Lippitz B. A simple dose gradient measurement tool to complement the conformity index. *Journal of neurosurgery.* 2006;105(Supplement):194-201.
24. Patton DR. A diversity index for quantifying habitat" edge". *Wildlife Society Bulletin (1973-2006).* 1975;3(4):171-173.
25. Kazhdan M, Simari P, McNutt T, et al. A shape relationship descriptor for radiation therapy planning. Paper presented at: International Conference on Medical Image Computing and Computer-Assisted Intervention2009.
26. Krayenbuehl J, Norton I, Studer G, Guckenberger M. Evaluation of an automated knowledge based treatment planning system for head and neck. *Radiation Oncology.* 2015;10(1):226.
27. Wu B, Kusters M, Kunze-busch M, et al. Cross-institutional knowledge-based planning (KBP) implementation and its performance comparison to Auto-Planning Engine (APE). *Radiotherapy and Oncology.* 2017;123(1):57-62.
28. Babier A, Mahmood R, McNiven AL, Diamant A, Chan TC. Knowledge‐based automated planning with three‐dimensional generative adversarial networks. *Medical Physics.* 2020;47(2):297-306.
29. Fan J, Wang J, Chen Z, Hu C, Zhang Z, Hu W. Automatic treatment planning based on three‐dimensional dose distribution predicted from deep learning technique. *Medical physics.* 2019;46(1):370-381.
30. Shen C, Nguyen D, Chen L, et al. Operating a treatment planning system using a deep‐reinforcement learning‐based virtual treatment planner for prostate cancer intensity‐modulated radiation therapy treatment planning. *Medical physics.* 2020.
31. Appenzoller LM, Michalski JM, Thorstad WL, Mutic S, Moore KL. Predicting dose‐volume histograms for organs‐at‐risk in IMRT planning. *Medical physics.* 2012;39(12):7446-7461.
32. Nguyen D, Long T, Jia X, et al. A feasibility study for predicting optimal radiation therapy

dose distributions of prostate cancer patients from patient anatomy using deep learning. *Scientific reports.* 2019;9(1):1-10.
33. Nguyen D, Jia X, Sher D, et al. 3D radiotherapy dose prediction on head and neck cancer patients with a hierarchically densely connected U-net deep learning architecture. *Physics in Medicine & Biology.* 2019;64(6):065020.